\documentclass[aps,prl,twocolumn,10pt,superscriptaddress,amsmath,amssymb]{revtex4-1}

\usepackage{graphicx}
\usepackage{subfigure}
\usepackage{epstopdf}
\usepackage{hyperref}
\usepackage{slashed}
\usepackage{bm}
\usepackage{feynmp}
\usepackage{color}

\newcommand{\unitvec}[1]{\hat{\mathbf{e}}_{#1}}
\newcommand{\nablab}{\boldsymbol{\nabla}}
\newcommand{\sigmab}{\boldsymbol{\sigma}}

\begin{document}

\title{Topological magnetic dipolar interaction and non-local electric magnetization control in topological insulator heterostructures}

\author{Stefan Rex}
\affiliation{Department of Physics, Norwegian University of Science and Technology, N-7491 Trondheim, Norway}
\author{Flavio S. Nogueira}
\affiliation{Institute for Theoretical Solid State Physics, IFW Dresden, PF 270116, 01171 Dresden, Germany}
\affiliation{Institut f{\"u}r Theoretische Physik III, Ruhr-Universit{\"a}t Bochum, Universit{\"a}tsstra{\ss}e 150, DE-44801 Bochum, Germany}
\author{Asle Sudb{\o}}
\affiliation{Department of Physics, Norwegian University of Science and Technology, N-7491 Trondheim, Norway}

\begin{abstract}
The magnetoelectric effect predicted in topological insulators makes heterostructures that combine magnetic materials and such insulators promising candidates for spintronics applications. Here, we theoretically consider a setup that exhibits two well-separated interfaces between a topological insulator and a ferromagnetic insulator. We show that there is a topological magnetic dipole-dipole interaction stemming from long-range Coulomb interactions. We analytically derive the magnetization dynamics at the two interfaces and discuss how the long-range coupling can be applied to non-locally induce the formation of a magnetic texture at one interface by suitably gating the other interface.
\end{abstract}

\maketitle

Topological insulators (TIs) represent a fascinating and novel state of matter, namely a combined bulk insulator and surface metal with the additional property that the gapless current-carrying surface states are protected from scattering by particle number conservation and time-reversal symmetry\cite{HasanKaneRMP, QiZhangRMP}. When TIs coexist with magnetic order, the magnetization opens a gap in the surface Dirac cone on the topological insulator. This leads to an anomalous quantum Hall effect with a half integer quantized conductance of $\sigma^0_{xy}=e^2/(2h)$ \cite{HasanKaneRMP, ZhengAndo2002}, and a topological magnetoelectric effect (TME) \cite{QiHZ2008} whereby an electric field induces a magnetic polarization in the same direction and vice versa. The latter can be understood from a field theoretic description of the Dirac fermions, which resembles axion electrodynamics\cite{Wilczek1987}. Namely, the TME is evoked by a contribution proportional to $\theta\boldsymbol{E\cdot B}$ in the Lagrangian, where $\theta$ is the axion field. This term is of a topological origin and quantized, as only the two values $\theta=0,\pi$ are allowed by time-reversal symmetry (TRS) of the TI bulk.

Even though conclusive direct evidence of the TME is still pending, the tantalizing idea of magnetization control by electric fields in a topologically protected way has lead to intense research. For instance, heterostructures of TIs and ferromagnets have recently attracted much attention as a highly promising platform for spintronics in both theory \cite{Garate10, Yokoyama10, Semenov12, Wickles12, Nogueira13, MSB13, Wang15, Ferreiros15} and experiment \cite{Moodera,kapitulnik-2013, Fan2014, Mellnik2014, WangPRL15,Moodera-bilayer-2-2015,Katmis}. The strong spin-orbit coupling required to invert the band structure in a TI enables strong spin-orbit torques, and the spin-momentum locking provided by the topological Dirac fermions offers unique possibilities for magnetization control by electrical currents. Envisioned devices based on the TME aim at, for instance, electrically controlled domain wall motion \cite{LossPRL12, Ferreiros14, Linder14}, Qubits\cite{FeL13}, and even indicate a route to topological transistors \cite{MooderaNatMat14}.

Importantly, the TME is a generic feature involving \emph{any} electric field that is present, not only external fields. The setup we propose in this Letter differs from most previous suggestions in that it takes into account the fluctuating electric field stemming from long-range Coulomb interaction and its impact on the magnetization dynamics in the presence of the TME \cite{NogueiraPRL, PRB16}. This is crucial, as the Coulomb interaction always will be present in a real system. For instance, electrostatic coupling between TI surfaces has been reported \cite{Fatemi2014}. We show that Coulomb interactions lead to a topological magnetic dipole-dipole interaction whereby it gives rise to a magnetic anisotropy. Furthermore, we suggest a spintronics nanodevice where this long-range interaction is exploited to couple two otherwise completely independent interfaces  TIs and 
ferromagnetic insulators (FMIs). A magnetic texture at one interface can then be switched on and off by applying a voltage at the other interface. Measuring this effect would not only serve as a clear evidence of the TME, since no other coupling mechanism exists in the system we consider. It might also inspire device architectures for electric magnetization control where the applied field and the desired response are locally separated in the device.

\begin{figure}[tb]
\includegraphics[width=0.9\columnwidth]{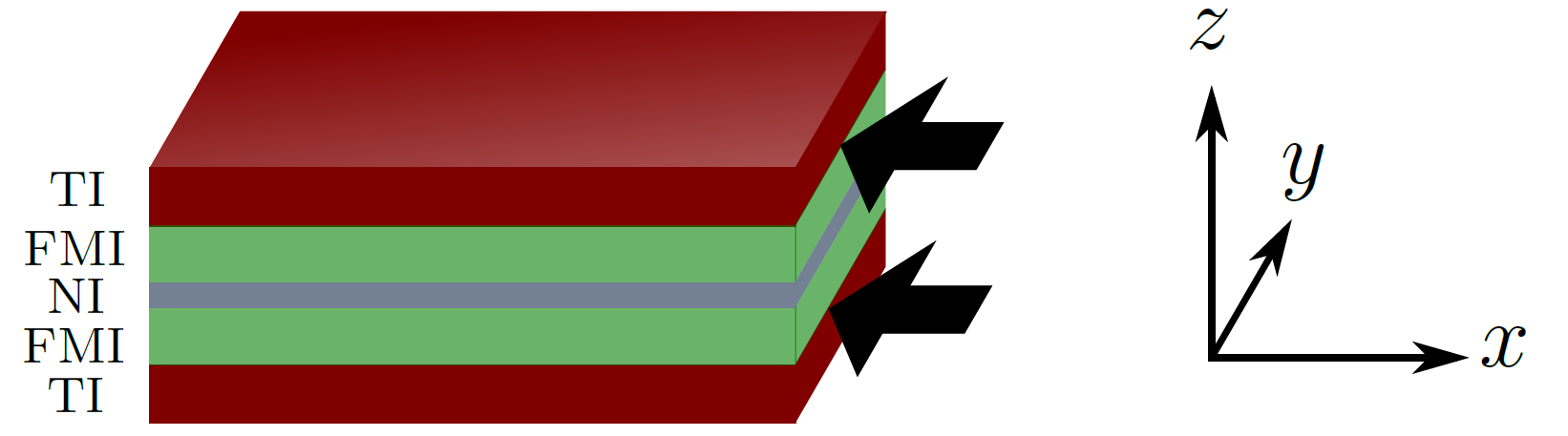}
\caption{The structure of the nanodevice: a top and a bottom TI layer sandwich two FMI layers that are separated by a nonmagnetic insulator (NI). At the two interfaces indicated by arrows, the magnetization opens a gap in the TI surface state dispersion, leading to the TME. The interfaces are well-separated and interact only via Coulomb interactions. The coordinate system is chosen such that the $z$ axis is pointing out of plane.}
\label{FigSetup}
\end{figure}

In Fig.~\ref{FigSetup}, we show a possible TI/FMI multilayer heterostructure for our approach featuring two parallel magnetoelectrically active interfaces. We emphasize that these two interfaces belong to different TI layers, and that these two layers are separated by FMI and nonmagnetic insulator layers in a way  that no electron hopping or direct magnetic coupling is present. In the following, we will first employ the framework of quantum field theory to obtain the effective Lagrangian of the system in the low-frequency regime after integrating out all quantum fluctuations, and show that it contains a topologically protected magnetic dipolar interaction. Then, we derive the Landau-Lifshitz equation (LLE) for magnetization dynamics. Finally, we argue how non-local electric magnetization control is possible with a suitable gate placement at one of the interfaces. We work in natural and Gaussian units, and assume that the TI and FMI layers are made of the same material, respectively, such that any material constants are the same at the two interfaces. We use the symbol $\nablab$ to denote the two-dimensional differential operator.

We begin by considering the Coulomb interaction, which plays a key role for our results. The well-known three-dimensional $r^{-1}$ potential is acting on charge carriers that are restricted to a plane. A two-dimensional Fourier transformation yields the intra-plane potential $2\pi e^2/|\mathbf{q}|$, and the inter-plane potential $2\pi e^2 \exp(-d|\mathbf{q}|)/|\mathbf{q}|$. Here, $e$ is the elementary charge, $\mathbf{q}$ the momentum in two dimensions, and $d$ denotes the distance between the two interfaces. 
The overlap integral of electron orbitals from different interfaces will be zero, since they belong to different TI bulks and are well-separated. The only contribution from the Coulomb interaction acting between the interfaces will thus be a density-density interaction, while exchange interactions vanish. To facilitate handling the two-particle interaction, we write it as a single-particle term by introducing the scalar Hubbard-Stratonovich fields $\varphi_i$ with units of an electric potential. With the operator $\rho_i(\mathbf{q})$ of electron density at interface $i=1,2$, the potential then has the two contributions $\sum_{i,\mathbf{q}}e\varphi_i(\mathbf{q})\rho_i(\mathbf{q})$ and $\frac{1}{2}\sum_{i,j,\mathbf{q}}\varphi_i(-\mathbf{q})B_{ij}(\mathbf{q})\varphi_j(\mathbf{q})$, where the matrix $B$ now contains the Coulomb-mediated coupling of the interfaces. The matrix entries can be derived from the intra- and inter-plane potential.

In the low-energy regime, the conduction electrons at the two interfaces can be described by a Lagrangian
\begin{equation}
\mathcal{L}_i = \Psi_i^\dagger\left[i\partial_t + iv_F(\sigmab\times\nablab)\cdot\unitvec{z} + e(\varphi_i+\phi_i) + J\sigmab\cdot\mathbf{n}_i\right]\Psi_i
,\end{equation}
with the second-quantized fermion operators $\Psi_i^\dagger, \Psi_i$. It contains the Dirac-cone dispersion proportional to the Fermi velocity $v_F$, which is typical for TI surface states. As explained above, the electric potential $\varphi_i$ from the Coulomb interaction enters. In addition, we allow for an externally applied electric field $\mathbf{E}_i=-\nablab\phi_i$. 
Proximity to an FMI layer induces a magnetization $\mathbf{n}_i$ at each interface that couples to the electron spin with a coupling strength $J$. The three Pauli matrices are included in the vector $\sigmab$. In the anomalous quantum Hall regime, the uniform mean-field magnetization will be orthogonal to the plane and give rise to a mass $m_\Psi=J\langle n_{1z}\rangle=J\langle n_{2z}\rangle$ of the fermion field, thus opening a gap in the Dirac cone. We assume that the Fermi level, 
$\epsilon_F$, 
lies inside the gap, either by doping or gating. Thus, there is no loss of generality in considering $\epsilon_F=0$. Furthermore, note that since 
 $\varphi_i$ are fluctuating fields, any nonzero $\epsilon_F$ can be absorbed into $\varphi_i$. The situation is different in the metallic regime where 
 the Fermi level lies outside the gap, in which case Friedel oscillations are expected to occur in the Coulomb interacting system 
\cite{Note-1}.

To account for the ferromagnetism of the bulk FMI layers, we add
\begin{equation}
\mathcal{L}_\text{FMI,i} = \mathbf{b}_i\cdot\partial_t\mathbf{n}_i - \frac{\kappa}{2}\left[(\nablab\mathbf{n}_i)^2+(\partial_z\mathbf{n}_i)^2\right]
,\end{equation}
where $(\partial_{\mathbf{n}_i}\times\mathbf{b}_i)\mathbf{n}_i^2=-\mathbf{n}_i$ and $\kappa>0$ 
is the coefficient of exchange energy. Recall that $\nablab$ is two-dimensional. We assume that 
the magnitude of the magnetization is fixed at constant temperature. In total, the model 
Lagrangian including Coulomb interactions is
\begin{eqnarray}
\mathcal{L}_\text{tot}(\mathbf{r}) &=& \sum_{i=1,2} \left[\mathcal{L}_i(\mathbf{r}) + \mathcal{L}_{\text{FMI},i}(\mathbf{r}) \right]\notag\\
&+&\frac{1}{4\pi}\sum_{i,j=1,2}\sum_{\mathbf{q}}\left[\nablab\varphi_i(\mathbf{r})\right]\cdot\int\text{d}^2r^\prime\frac{e^{i(\mathbf{r}-\mathbf{r}^\prime) \cdot \mathbf{q}}}{|\mathbf{q}|\left(1-e^{-2|\mathbf{q}|d}\right)}\notag\\
&&\qquad\left(\delta_{ij}-(1-\delta_{ij})e^{-|\mathbf{q}|d}\right)\left[\nablab^\prime\varphi_j(\mathbf{r}^\prime)\right]
.\end{eqnarray}
Since our analysis will mainly focus the magnetoelectric dynamics of the magnetization at the interface, the physics is effectively two-dimensional. 
Furthermore, we will perform the calculations at zero temperature. However, it must be noted that the results obtained here are of relevance 
for finite temperature analyzes, provided $T\ll m_\Psi$. Due to the proximity coupling to the Dirac fermions at the interface, the system 
overcomes the Mermin-Wagner theorem so that a finite Curie temperature exists. To see this it is enough to consider a mean-field theory where 
the dynamics of the magnetization is simply given by, 
$\partial_t{\bf n}={\bf n}\times\left({\bf H}_{\rm eff}+J\langle\Psi^\dagger\sigmab\Psi\rangle\right)$,  where 
where ${\bf H}_{\rm eff}=-\delta{\cal H}_{\rm FM}/\delta{\bf n}$, with ${\cal H}_{\rm FM}$ being the Hamiltonian of the ferromagnet. In this case 
it is easy to show that the magnon spectrum is given by $\omega(q)=m_\Psi[(\kappa/J)q^2+1]$, which clearly does not have any infrared 
singularity. 

\begin{figure}[tb]
\includegraphics{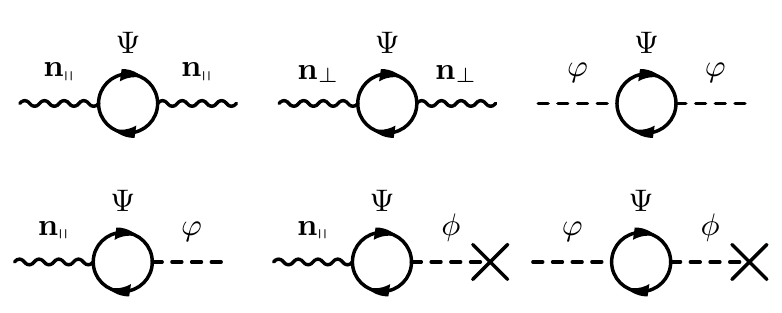}
\caption{The topologically distinct one-loop diagrams of vacu\-um polarization that contribute to the effective action upon integrating out the fermions. Here, $\mathbf{n}_\shortparallel$ and $\mathbf{n}_\perp$ denote the magnetization fluctuations in-plane and out-of-plane, respectively, $\Psi$ is the fermion field, $\varphi$ is the fluctuating Coulomb potential, and $\phi$ the electric potential that is fixed by the externally applied field (indicated with a cross). Diagrams that mix in-plane and out-of-plane fluctuations vanish, and we skipped the $\phi$-$\phi$ diagram, which yields a constant. Magnetoelectric effects are due to the first two diagrams in the second line.}
\label{FigDiagrams}
\end{figure}

We now integrate out the fermions, keeping only leading order terms. More precisely, we consider the one-loop diagrams of vacuum polarization. The relevant topologically distinct diagrams are shown in Fig.~\ref{FigDiagrams}. The breaking of TRS by the magnetization generates a Chern-Simons (CS) term in the resulting action\cite{QiHZ2008}
that leads to the TME $\sim(\mathbf{n}\times\partial_t\mathbf{n})$ and a Berry phase.
Since the two interfaces so far are decoupled in the fermion sector, to this point the calculation is identical for $i=1$ and $2$. The field-theoretic treatment of a single interface can be found in Ref.~\cite{PRB16}. Next, we proceed to integrate out the Hubbard-Stratonovich fields
to unravel the effective magnetic interaction. The Lagrangian for the fields 
$\mathbf{n}_i(\mathbf{r})$ is then given by
\begin{eqnarray}
\label{Eq:Leff}
&&\mathcal{L}(\mathbf{r}) =
\sum_{i=1,2}\Bigg[\mathcal{L}_{\text{FMI},i}(\mathbf{r})-\frac{\sigma_{xy}}{2v_F^2}
\left(\mathbf{n}_i(\mathbf{r})\times\partial_t\mathbf{n}_i(\mathbf{r})\right)\cdot\unitvec{z}
\nonumber\\
&&-\frac{NJ^2}{24\pi m_\Psi}\left[(\nablab\cdot\mathbf{n}_i(\mathbf{r}))^2+(\nablab n_{i,z}(\mathbf{r}))^2\right]
\nonumber\\
&&-\frac{NJ^2m_\Psi}{2\pi v_F^2}n_{i,z}^2(\mathbf{r})
+ \frac{NJm_\Psi^2}{\pi v_F^2}n_{i,z}(\mathbf{r})
-\frac{\sigma_{xy}e}{Jv_F}\mathbf{n}_i(\mathbf{r})\cdot\mathbf{E}_i(\mathbf{r})
\Bigg]
\nonumber\\
&&+\sum_{i,j=1,2}\frac{\sigma_{xy}e}{2Jv_F}\left[\nablab\cdot\mathbf{n}_i(\mathbf{r})\right]
\int\frac{\text{d}^2r^\prime\rho_j(\mathbf{r}^\prime)}{\sqrt{(\mathbf{r}-\mathbf{r}^\prime)^2+(1-\delta_{ij})d^2}}
\end{eqnarray}
Here, $\sigma_{xy} = \sigma^0_{xy}NJ^2/e^2$ is the induced Hall conductivity, assuming $N$ electron orbital degrees of freedom. We have neglected several further terms that are constant or contain time-derivatives that are not of leading order in the low-frequency regime. 
Although we have not included any intrinsic axial anisotropy in the Lagrangian of the FM proximate to the TI, we note that 
such an anisotropy has been dynamically generated by Dirac fermions quantum fluctuations, in the form of a term $\sim n_{iz}^2$. 
Thus, if an intrinsic axial anisotropy is already present in the FM, the TI surface states will necessarily enhance it.  
In the last term, $\rho_j$ denotes the effective charge density which appears in the presence of in-plane divergences of the magnetization and the applied electric field. This is because the electric charge at a TI/FMI interface coincides with the magnetic charge\cite{NomNag2010}. The charge density is given by
\begin{equation}
\rho_i = \frac{\sigma_{xy}e}{2v_FJ}\nablab\cdot\mathbf{n}_i - \frac{Ne^2}{24\pi m_\Psi}\nablab\cdot\mathbf{E}_i
\label{EqChargeDensity}
,\end{equation}
where the first term is of topological origin. Remarkably, the contribution from this topological term to Eq.~\eqref{Eq:Leff} can be rewritten by partial integrations over both $\mathbf{r}$ and $\mathbf{r}^\prime$ to read
\begin{eqnarray}
\label{Eq:dipolar}
\mathcal{L}_\text{dipolar}(\mathbf{r}) &=& -\left(\frac{\sigma_{xy}e}{2Jv_F}\right)^2\!\!\!\!\!\sum_{i,j=1,2}\int \frac{\text{d}^2r^\prime}{\left[(\mathbf{r}-\mathbf{r}^\prime)^2+(1-\delta_{ij})d^2\right]^{3/2}}
\nonumber\\
&\times&\Bigg\{3\frac{\left[\mathbf{n}_i^\parallel(\mathbf{r})\cdot(\mathbf{r}-\mathbf{r}^\prime)\right]\left[\mathbf{n}_j^\parallel(\mathbf{r}^\prime)\cdot(\mathbf{r}-\mathbf{r}^\prime)\right]}{(\mathbf{r}-\mathbf{r}^\prime)^2+(1-\delta_{ij})d^2}
\nonumber\\
&-&\mathbf{n}_i^\parallel(\mathbf{r})\cdot\mathbf{n}_j^\parallel(\mathbf{r}^\prime)\Bigg\},
\end{eqnarray}
where ${\bf n}_i^\parallel=(n_{ix},n_{iy},0)$. 
Thus, we have found a magnetic dipole-dipole interaction having an intrinsic topological topological origin.   We note from the 
effective magnetic field, ${\bf H}_{\rm eff}^{(i)}=-\partial{\cal H}/\partial{\bf n}_i$, that it is the dipolar interaction that connects 
the two interfaces via the in-plane magnetization. Here ${\cal H}$ is the Hamiltonian associated to the effective Lagrangian 
(\ref{Eq:Leff}), i.e., by removing the Berry phase terms.  

Typically, dipolar interactions generate a magnetic anisotropy, turning the susceptibility non-diagonal. Indeed, it is easy to see from Eqs. 
(\ref{Eq:Leff}) and (\ref*{Eq:dipolar}) that the susceptibilities for the spin-wave modes 
in the interfaces and across them feature transverse and longitudinal components, and have the form, 
$\chi_{\alpha\beta}^{ii}(\omega,{\bf q})=\chi_T^{ii}(\omega,q)(\delta_{\alpha\beta}-q_\alpha q_\beta/q^2)+\chi_L^{ii}(\omega,q)q_\alpha q_\beta/q^2$, 
$\chi_{\alpha\beta}^{12}(\omega,{\bf q})=\chi_{\alpha\beta}^{21}(\omega,{\bf q})=\chi_T^{12}(\omega,q)(\delta_{\alpha\beta}-q_\alpha q_\beta/q^2)+\chi_L^{12}(\omega,q)q_\alpha q_\beta/q^2$, where $\alpha,\beta=x,y$, 
and $\chi_{zz}^{ii}(\omega,q)$ describes the gapped, longitudinal (in field space), mode. The spin-wave mode across the interfaces decays exponentially 
with the thickness in momentum space. Moreover, there is no longitudinal field mode propagating between the interfaces.   
Dipolar interactions are normally considerably smaller than 
exchange interactions. However, they are known to be as large as exchange 
interactions in some ferromagnetic insulators, such as Europium 
monochalcogenides \cite{Kasuya-PRB-1973}. The dipolar 
interaction (\ref{Eq:dipolar})  
is quantized due to the TME. An estimate can be given based on recent experiments on Bi$_2$Se$_3$-EuS heterostructures 
\cite{Moodera,Katmis}. Using $\hbar v_F=2.17$ eV $\cdot$ \AA ~and assuming that $J\approx 90$ meV, we estimate a dipolar 
interaction roughly having a strength $\sim 1$ meV.   
Note how the prefactor in Eq. (\ref{Eq:dipolar})
is independent of the fermionic gap $m_\Psi$. Thus, the topologically induced
dipolar term  is expected to play a role also above the 
Curie temperature of the system. In principle, the anisotropy in 
the susceptibility can be probed in the static limit 
via polarized neutron scattering techniques, similarly to the one 
used in Ref. \cite{Boeni-1991} to probe the dynamics of 
longitudinal and transverse fluctuations in EuS.  Since Eq. (\ref{Eq:dipolar}) involves only the planar components of the 
magnetization, it is particularly sensitive to polarized neutron reflectometry (PNR) experiments, since PNR only measures 
the in-plane components of the magnetization.  
In the context of TI heterostructures, PNR has 
recently been successfully 
used to probe the magnetization for a wide range of temperatures near the 
interface between Bi$_2$Se$_3$ and EuS in a TI/FMI 
bilayer structure \cite{Katmis}. The same method can in principle be 
used to find evidence of a dipolar magnetic anisotropy arising from TME.      

From Eq.~\eqref{Eq:Leff}, we derive the LLE at interface $i$,
\begin{equation}
\left(\frac{\mathbf{n}_i}{\mathbf{n}_i^2}+\frac{\sigma_{xy}}{v_F^2}\unitvec{z}\right)\times\partial_t\mathbf{n}_i = \mathbf{d}_{\mathbf{n},i} + \mathbf{d}_{\mathbf{E},i} + \mathbf{d}_{\text{Cou},i}
\end{equation}
which describes precession around an effective field $\mathbf{d}_i$.
The second term inside the parentheses stems from the additional Berry phase generated by the CS term.
The effective field consists of three contributions: $\mathbf{d}_{\mathbf{n},i}$ describes the local spin dynamics and is given by 
\begin{eqnarray}
\mathbf{d}_{\mathbf{n},i} &=& \rho_\text{s}\cdot\left(\nablab^2\mathbf{n}_i\right) + \frac{NJ^2}{12\pi m_\Psi}\nablab(\nablab\cdot\mathbf{n}_i)\notag\\
&&{}+ \frac{NJm_\Psi}{\pi v_F^2}\left(Jn_{i,z}-m_\Psi\right)\unitvec{z}
,\end{eqnarray}
with the spin-stiffness matrix
$\rho_\text{s} = \text{diag}\left(\kappa,\kappa,\kappa+NJ^2/(12\pi m_\Psi)\right)$.
The last line reflects the dynamically generated axial anisotropy and leads to precession around the out-of-plane axis even if the magnetization is uniform.
The vectors $\mathbf{d}_{\mathbf{E},i}=\sigma_{xy}e/(Jv_F)\mathbf{E}_i$ and $\mathbf{d}_{\text{Cou},i}=\sigma_{xy}e/(Jv_F)\mathbf{E}_{\text{Cou},i}$ are due to the TME involving the external field and the Coulomb field of the charge density Eq.~\eqref{EqChargeDensity}, respectively. Thus, $\mathbf{d}_{\text{Cou},i}$ is non-local and contains both in-plane and inter-plane interactions.
The Coulomb field at interface $i$ is given by
\begin{equation}
\mathbf{E}_{\text{Cou},i}(\mathbf{r}) = -\sum_{j=1,2}\int\!\text{d}^2r^\prime\,\frac{(\mathbf{r}-\mathbf{r}^\prime)\rho_j(\mathbf{r}^\prime)}{[(\mathbf{r}-\mathbf{r}^\prime)^2+(1-\delta_{ij})d^2]^{3/2}}
\label{EqCoulombField}
.\end{equation}
Eqs.~(\ref{EqChargeDensity}) and \eqref{EqCoulombField} describe the effective non-local interaction between magnetic moments in the system. We find that a charge density at \emph{one} interface leads to a net in-plane magnetic texture at \emph{both} interfaces.

\begin{figure}[tb]
\begin{center}
\includegraphics[width=0.7\columnwidth]{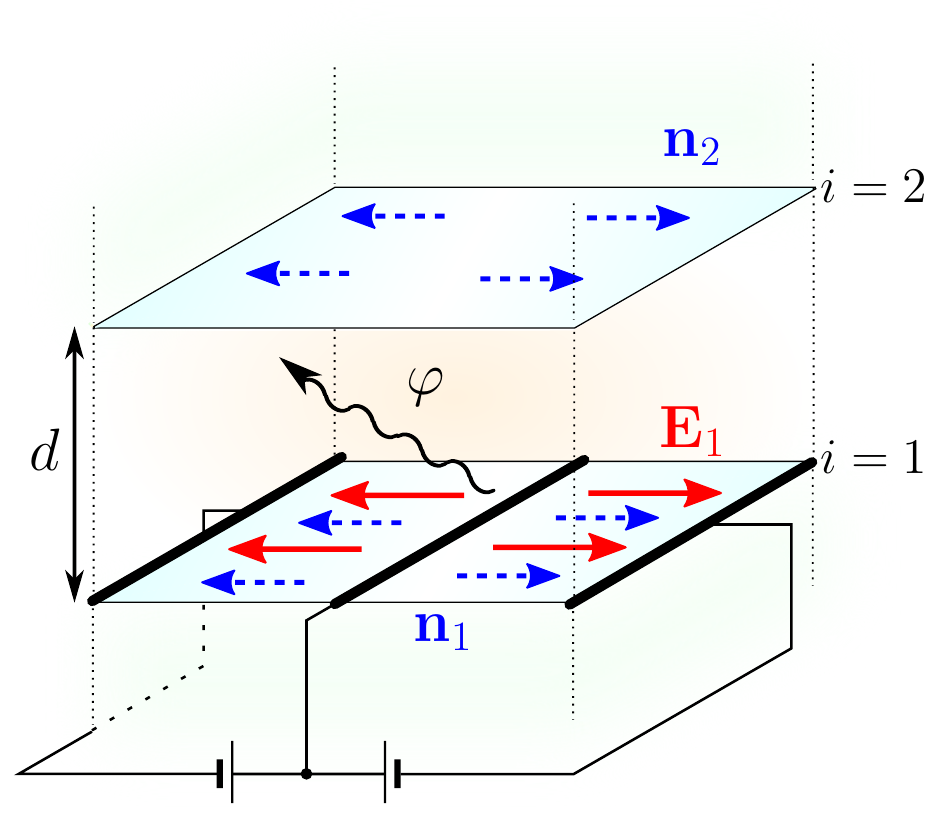}
\caption{Schematic depiction of the mechanism for a specific gate placement: At interface 1, a voltage between the middle and edge gates (bold black lines) leads to an electric field $\mathbf{E}_1$ (solid red arrows) with an in-plane divergence. By the TME, the in-plane component of the magnetization (dashed blue arrows) $\mathbf{n}_1$ alignes with the field, resulting in a charged texture that gives rise to a Coulomb potential $\varphi$. The Coulomb field causes the magnetization $\mathbf{n}_2$ at interface 2 to develop a magnetic texture as well.}
\label{FigSketch}
\end{center}
\end{figure}

Based on this topological coupling mechanism, we propose a spintronics device for nonlocal electric magnetization control, where we adopt the following strategy: one of the magneto-electrically active interfaces is gated such that the electric field will have an in-plane divergence. By the TME, the magnetization at the same interface will develop a net in-plane component that is aligned with the field. Thus, a charge density according to Eq.~\eqref{EqChargeDensity} is induced and creates a field, cf. Eq.~\eqref{EqCoulombField}, that finally causes a magnetic texture to emerge at the other interface. A specific gate geometry is shown in Fig.~\ref{FigSketch}, where we consider the impact of an applied electric field at $i=1$ on the magnetic texture at $i=2$. We place three gates, where the two outer ones lie at the same potential, and a voltage is applied between them and the middle gate such that the electric field will have opposite orientation in the two half-planes. Consequently, a charge density emerges along the middle gate, where $\nablab\cdot\mathbf{E}_1$ and $\nablab\cdot\mathbf{n}_1$ become large. Due to the Coulomb field, the net magnetization at interface $2$ will develop an opposite in-plane component in the two half-planes. Thus, we obtain a magnetic texture without any local manipulations. This texture can be switched on and off by means of the voltage applied at the first interface. We note that in principle any setup where the applied field has a divergence would work.

Besides applications for electric magnetization control, measuring the magnetic texture at interface $2$ would also be an intriguing demonstration of the TME. The inter-plane coupling mechanism is topologically protected. Namely, the TME that translates the diverging field into a magnetic charge density at interface $1$, the correspondence of magnetic and electric charge, and the TME with the Coulomb field at interface $2$ are all topologically protected. Furthermore, the device is constructed in such a way that other long-range interactions are excluded. One could think of a seemingly simpler heterostructure than the one shown in Fig.~\ref{FigSetup}, where a single TI layer is coated with FMI layers on both sides, such that the active interfaces are opposite surfaces of the same TI bulk. In that case, however, the electric field could, at least close to the sample edges, directly leak around the topological side surfaces onto the other interface and interfere with the magnetization there, circumventing the desired long-range coupling of purely topological origin. The nonmagnetic insulator layer in our setup also prevents spin waves from traveling from one interface to the other.

In conclusion, we have analytically derived a topological magnetic dipole-dipole interaction that emerges from long-range Coulomb interactions in the presence of the TME. It generates a magnetic anisotropy that could, e.g., be probed by PNR. 
We presented analytical results for the magnetization dynamics in a heterostructure with two well-separated parallel TI/FMI interfaces and demonstrated that the long-range interactions enable non-local electric control of a magnetic texture at one interface by applying a voltage at the other interface. We believe that these results are experimentally accessible with the existing technology.

\textsl{Acknowledgements.} S. R. and A. S. acknowledge support by the Norwegian 
Research Council, grants 205591/V20 and 216700/F20. F. S. N. would like to thank 
the Collaborative Research Center SFB 1143 ``Correlated Magnetism: From Frustration 
to Topology" for the financial support.



\begin{thebibliography}{-}
\bibitem{Note-1}  Absorbing $\epsilon_F$ into $\varphi_i$ does not make the Fermi level irrelevant, since in this case the Fermi momentum is given by 
$k_F^2=\sqrt{\epsilon_F^2-m_\Psi^2}$.
\end{thebibliography}
\end{document}